\documentclass[
  aps,
  prd,
  twocolumn,
  superscriptaddress,
  nofootinbib,
  longbibliography
]{revtex4-2}

\usepackage{amsmath,amssymb,amsfonts}
\usepackage{graphicx}
\usepackage{bm}
\usepackage{booktabs}
\usepackage{multirow}
\usepackage{float}
\usepackage{xcolor}
\usepackage{adjustbox}
\usepackage[mathscr]{euscript}
\providecommand{\mathpzc}[1]{\EuScript{#1}}
\allowdisplaybreaks[2]

\usepackage[colorlinks=true,linkcolor=blue,citecolor=blue,urlcolor=blue]{hyperref}

\newcommand{\Tr}{\mathrm{Tr}}
\newcommand{\dd}{\mathrm{d}}
\newcommand{\cO}{\mathcal{O}}
\newcommand{\cP}{\mathcal{P}}

\newcommand{\para}{\parallel}
\newcommand{\perpdir}{\perp}

\begin{document}

\title{Distribution amplitudes of vector and axial--vector mesons in a nonperturbatively improved symmetry-preserving framework}

\author{Z.-N.\ Xu%
$^{\href{https://orcid.org/0000-0002-9104-9680}{\textcolor[rgb]{0.00,1,0.00}{\sf ID}}}$}
\affiliation{Dpto.\ Ciencias Integradas, Centro de Estudios Avanzados en Fis., Mat.\ y Comp.,
Fac.\ Ciencias Experimentales, \href{https://ror.org/03a1kt624}{Universidad de Huelva},
E-21071 Huelva, Spain}

\author{Z.-Q.\ Yao%
$\,^{\href{https://orcid.org/0000-0002-9621-6994}{\textcolor[rgb]{0.00,1,0.00}{\sf ID}}}$}
\email{z.yao@hzdr.de}
\affiliation{\href{https://ror.org/01zy2cs03}{Helmholtz-Zentrum Dresden-Rossendorf},
Bautzner Landstra{\ss}e 400, D-01328 Dresden, Germany}

\author{P.\ Cheng$\,^{\href{https://orcid.org/0000-0002-6410-9465}{\textcolor[rgb]{0.00,1.00,0.00}{\sf ID}}}$}
\affiliation{Department of Physics, \href{https://ror.org/05fsfvw79}{Anhui Normal University}, Wuhu, Anhui 24100, China}

\author{K. Raya%
$^{\href{https://orcid.org/0000-0001-8225-5821}{\textcolor[rgb]{0.00,1,0.00}{\sf ID}}}$}
\email{khepani.raya@dci.uhu.es}
\affiliation{Dpto.\ Ciencias Integradas, Centro de Estudios Avanzados en Fis., Mat.\ y Comp.,
Fac.\ Ciencias Experimentales, \href{https://ror.org/03a1kt624}{Universidad de Huelva},
E-21071 Huelva, Spain}


\begin{abstract}
Using continuum Schwinger-function methods with a nonperturbatively
improved, symmetry-preserving kernel, we deliver predictions for the
leading-twist light-front distribution amplitudes (DAs) of the
$\rho$, $K^\ast$, $a_1(1260)$, $b_1(1235)$, and the unmixed strange
partners of the $1^{++}$ and $1^{+-}$ axial-vector (AV) channels,
reconstructed from Mellin moments of the associated
Bethe--Salpeter wave functions. For vector mesons, polarisation
barely affects longitudinal momentum sharing: the longitudinal and
transverse DAs are nearly degenerate, and both narrower than the
asymptotic distribution in the second-moment sense. The AV sector is
different in kind. Charge conjugation compels one projection --
interchanged between the $1^{++}$ and $1^{+-}$ channels -- to vanish
at $x=1/2$ and change sign; breaking $SU_F(3)$ symmetry removes this
protection, whereupon the zeroth moments become nonzero and the nodes
shift from the midpoint. Under a common weighted normalisation, the
$1^{+-}$ zeroth moment is $1.63$ times that of the $1^{++}$ channel,
and the profile distortion follows the same pattern. A coupling
forbidden by charge conjugation in the symmetric limit,
$f_{b_1}=0$, becomes $f_{K_1^{+-}}=0.019\,$GeV in the strange channel:
an independent measure of the same symmetry breaking, obtained from a
current matrix element rather than from the DA reconstruction. What distinguishes the two
sectors is thus a symmetry-enforced zero, not the size of the flavour
asymmetry.
\end{abstract}

\keywords{
Distribution amplitudes;
axial--vector mesons;
Dyson--Schwinger equations;
Bethe--Salpeter equation;
symmetry-preserving truncation
}

\maketitle

\section{Introduction}
\label{sec:intro}

Parton distribution amplitudes (DAs) describe how a hadron's valence
constituents share longitudinal momentum on the light front. They are indispensable nonperturbative inputs to QCD descriptions of hard exclusive
processes\,\cite{Efremov:1979qk,Lepage:1979zb,Lepage:1980fj}; and
they link such observables to Poincar\'e-covariant bound-state
dynamics\,\cite{Raya:2024ejx,Roberts:2020udq,Ding:2022ows,Lu:2023yna,Binosi:2018rht}.
For ground-state pseudoscalar and vector mesons, the leading-twist
DAs of light systems are broad, concave and positive definite;
they narrow as the valence-quark masses increase, and unequal
masses skew
them\,\cite{Serna:2020txe,Xu:2025hjf,Cui:2020tdf,Cheng:2020vwr,Chang:2025lrc,RQCD:2019osh,Zhang:2020gaj,LatticeParton:2022zqc}.
These patterns are a direct expression of the competition between
emergent hadron mass (EHM) and the Higgs-generated current-quark
masses.

Axial--vector (AV) mesons are qualitatively
different\,\cite{Chang:2010hb,Chang:2011ei,Xu:2022kng,Qin:2020jig,Ferreira:2026gbe,Hagel:2025ngi}.
Charge conjugation acts differently on their longitudinal and
transverse light-front projections: in the flavour-symmetric limit,
the leading-twist DA of the $1^{++}$ channel is symmetric under
$x\leftrightarrow1-x$ in the longitudinal projection and
antisymmetric in the transverse, a pattern reversed for the
$1^{+-}$ channel\,\cite{Yang:2007zt}. The antisymmetric projection changes sign at $x=1/2$ and has a
vanishing zeroth moment. No
analogous symmetry-enforced sign change is admitted by leading-twist
pseudoscalar and vector DAs. AV DAs are therefore keen probes of
charge-conjugation constraints and spin-dependent bound-state
dynamics.

This distinction is especially useful for strange AV mesons, where the
current-quark masses are unequal. In the kaon, flavour-symmetry breaking
skews an already nonzero, unit-normalised DA, and continuum and lattice
calculations place $|\langle\xi\rangle_K|$ at only a few per
cent\,\cite{Shi:2014uwa,Arthur:2010xf,RQCD:2019osh}. For a
sign-changing AV projection, by contrast, the zeroth moment vanishes in
the flavour-symmetric limit. Once the quark masses become unequal, this
protection is removed. In the calculation below, the resulting mass
splitting generates a nonzero zeroth moment and shifts the node away
from $x=1/2$. The corresponding decay constant also vanishes in the
symmetric limit and provides a complementary, although correlated,
measure of the same symmetry breaking.

In spectroscopic language, the $1^{++}$ and $1^{+-}$ channels are
commonly associated with the spin-orbital configurations $^3P_1$ and
$^1P_1$. In a Poincar\'e-covariant treatment, such labels are mere
shorthand: the BSA contains all Dirac covariants allowed by the quantum
numbers, and its light-front projections expose interference among them
that no rest-frame classification anticipates. The same covariants make
AV mesons one of the channels in which leading-order rainbow--ladder
(RL) truncation is least reliable, since it does not fully account for
the spin--orbit interactions generated by
EHM\,\cite{Fischer:2009jm,Chang:2009zb,Qin:2011xq,Chang:2011ei}.
Quantitative predictions therefore call for a symmetry-preserving
kernel that goes beyond RL.

Continuum Schwinger-function methods (CSMs), viz.,
Dyson--Schwinger and Bethe--Salpeter equations, provide a
symmetry-preserving framework in which hadron spectroscopy and
structure can be treated self-consistently; see, e.g.,
Refs.\,\cite{Roberts:1994dr,Maris:2003vk,Holl:2005st,Qin:2020jig,Xu:2022kng,Yao:2024ixu,Yao:2024uej,Xu:2023izo,Miramontes:2025imd,Eichmann:2026ttr}.
Within this framework, the same Bethe--Salpeter wave functions (BSWFs)
that determine meson masses and decay constants also yield Mellin
moments of DAs through suitable light-front projections. Moreover, a
symmetry-consistent Bethe--Salpeter kernel can be constructed from the
gap equation, so that one may move beyond RL while preserving the
relevant vector and axial--vector Ward--Green--Takahashi
identities\,\cite{Munczek:1994zz,Bender:1996bb,Qin:2020jig,Xu:2022kng}.

Hitherto, this approach has been widely employed to pseudoscalar and vector
mesons\,\cite{Ding:2015rkn,Binosi:2018rht,Serna:2020txe,Xu:2025hjf,Xu:2026mds},
whereas AV meson DAs have received much less attention. They have been
studied using QCD sum rules\,\cite{Yang:2007zt} and, for the $a_1$, in an early light-cone
quark-model calculation\,\cite{Ji:1992yf}. The results reported here are, to our knowledge, the first continuum
predictions for axial--vector meson distribution amplitudes.

Herein, we compute the leading-twist DAs of the $\rho$, $K^\ast$,
$a_1(1260)$, $b_1(1235)$, and the strange partners of the $1^{++}$
and $1^{+-}$ channels. Mellin moments are obtained directly from the
associated BSWFs; and the pointwise DAs are reconstructed using a
compact exponential form for the positive-definite distributions and
a Gegenbauer expansion for the sign-changing ones. Charge
conjugation not being a good quantum number for strange mesons,
$K_1^{++}$ and $K_1^{+-}$ here denote the unmixed strange partners
connected continuously to the flavour-symmetric channels, viz.\ not
the physical $K_1(1270)$ and $K_1(1400)$ states. In the vector
sector, the longitudinal and transverse DAs are found to be nearly
identical. In the AV sector, the calculation respects the symmetry
interchange between the two projections and reveals, in the $1^{+-}$
channel, a flavour-breaking response roughly twice that in the
$1^{++}$ channel; the zeroth moments, the profile distortions and
the decay constants each deliver this ordering independently.

The manuscript is organised as follows: Sec.~\ref{sec:da_defs} defines
the vector and AV DAs and their Mellin moments; Sec.~\ref{sec:gap_bse}
describes the symmetry-preserving continuum framework used to calculate
the BSWFs; Sec.~\ref{sec:reconstruct} presents the moments and pointwise
reconstructions; and Sec.~\ref{sec:concl} summarises our conclusions and
outlook.

\section{Parton distribution amplitudes}
\label{sec:da_defs}

We define the leading-twist DAs of vector and AV mesons through light-front projections of the corresponding Poincaré-covariant BSWFs. Let $P$ be the meson's total momentum and 
$n$ a light-like four-vector satisfying $n^2=0$ and $n\cdot P = -m_H$, where $m_H$ is the meson mass. For a quark--antiquark system, we write:
\begin{equation}
k_\pm = k \pm P/2,
\label{eq:kpm}
\end{equation}
and define the BSWF as follows:
\begin{equation}
\chi_\nu(k,P)
=
S_{f_1}(k_+)\,\Gamma_\nu(k,P)\,S_{f_2}(k_-)\,.
\label{eq:bswf}
\end{equation}
Here $S_{f_i}$ denotes the $f_i$-flavored dressed-quark propagator, while
$\Gamma_\nu$ is the meson BSA, with full covariant decomposition provided in Appendix\,\ref{app:chebyshev}.
Throughout, $f_1$ labels the heavier valence quark; hence, in the strange
systems, $x$ is the light-front momentum fraction carried by the $s$ quark.

The light-front momentum fraction carried by the valence quark is defined as:
\begin{equation}
x=\frac{n\cdot k_+}{n\cdot P},
\qquad 0\leq x\leq 1\,.
\label{eq:lfx}
\end{equation}
Accordingly, we introduce:
\begin{equation}
\delta_x(k_+)
:=
\delta\left(x-\frac{n\cdot k_+}{n\cdot P}\right)\,.
\label{eq:delta_x}
\end{equation}
We also introduce the light-like conjugate vector $\bar n$ and the
transverse projector
\begin{equation}
\begin{aligned}
n^2&=\bar n^2=0,\qquad n\cdot\bar n=-1,\\
\cO^\perp_{\rho\nu}
&=\delta_{\rho\nu}+n_\rho\bar n_\nu+\bar n_\rho n_\nu.
\end{aligned}
\label{eq:nbar}
\end{equation}

Thus, for $J^P=1^\pm$ mesons, the longitudinal  and transverse leading-twist DAs ($\varphi^\parallel$ and $\varphi^\perp$, respectively) are obtained from the following light-front projections:
\begin{align}
f_{1^\pm}^{\para}\,\varphi_{1^\pm}^{\para}(x)
&=
N_c Z_2 M_{1^\pm}\,
\Tr_D \int^\Lambda \frac{\dd^4k}{(2\pi)^4}\,
\delta_x(k_+)
\frac{1}{(n\cdot P)^2}
\nonumber\\
&\quad\times
\cP_{1^\pm}\,
\gamma\!\cdot\! n\, n_\nu\,
\chi_\nu(k,P)\,,
\label{eq:DA_Vpar}
\\[1.2ex]
f_{1^-}^{\perpdir}\,\varphi_{1^\pm}^{\perpdir}(x)
&=
-\frac{N_c Z_T}{2}\,
\Tr_D \int^\Lambda \frac{\dd^4k}{(2\pi)^4}\,
\delta_x(k_+)
\frac{1}{n\cdot P}
\nonumber\\
&\quad\times
\cP_{1^\pm}\,
n_\mu\sigma_{\mu\rho}\,
\cO^\perp_{\rho\nu}\,
\chi_\nu(k,P)\,.
\label{eq:DA_Vperp}
\end{align}
Here $N_c=3$, $Z_2$ is the quark wave-function renormalisation constant, $Z_T$ is the tensor-current renormalisation constant, and
$f_{1^\pm}^{\para,\perpdir}$ are the longitudinal and transverse vector-meson decay constants. The  operators $\cP_{1^\pm}$ are defined as follows:
\begin{equation}
    \label{eq:projOpe}
    \cP_{1^-}=1\,,\,\cP_{1^+}=\gamma_5\,,
\end{equation}
where the latter is a reflection of the intrinsic opposite parity between vector and AV mesons. The Mellin moments of the distributions are defined as usual,
\begin{equation}
\langle x^m\rangle =
\int_0^1 \dd x\, x^m\,\varphi(x)\,.
\label{eq:mellin_general}
\end{equation}
Therefore, using Eqs.~\eqref{eq:DA_Vpar} and
\eqref{eq:DA_Vperp}, we find:
\begin{align}
\langle x^m\rangle_{\para}^{1^\pm}
&=
\frac{N_c Z_2 M_{1^\pm}}{f_{1^\pm}^{\para}}\,
\Tr_D \int^{\Lambda} \frac{\dd^4k}{(2\pi)^4}\,
\frac{(n\!\cdot\!k_+)^{m}}{(n\!\cdot\!P)^{m+2}}
\nonumber\\
&\quad\times
\cP_{1^\pm}\,
\gamma\!\cdot\!n\,n_\nu\,
\chi_\nu(k,P)\,,
\label{eq:mellinVpar}
\\[1.2ex]
\langle x^m\rangle_{\perpdir}^{1^\pm}
&=
-\frac{N_c Z_T}{2 f_{1^\pm}^{\perpdir}}\,
\Tr_D \int^{\Lambda} \frac{\dd^4k}{(2\pi)^4}\,
\frac{(n\!\cdot\!k_+)^{m}}{(n\!\cdot\!P)^{m+1}}
\nonumber\\
&\quad\times
\cP_{1^\pm}\,
n_\mu\sigma_{\mu\rho}\,
\cO^\perp_{\rho\nu}\,
\chi_\nu(k,P)\,.
\label{eq:mellinVperp}
\end{align}

It is worth noting that in flavour-symmetric systems, charge conjugation fixes the behaviour of
the DAs under the exchange $x\leftrightarrow 1-x$. For vector mesons with
$J^{PC}=1^{--}$, both leading-twist DAs are symmetric:
\begin{equation}
\varphi_{1^-}^{\para}(x)=\varphi_{1^-}^{\para}(1-x),
\qquad
\varphi_{1^-}^{\perpdir}(x)=\varphi_{1^-}^{\perpdir}(1-x).
\label{eq:sym_vector}
\end{equation}
For AV mesons, the symmetry pattern depends on the
charge-conjugation quantum number and on the light-front projection:
\begin{equation}
\begin{alignedat}{2}
1^{++}\ ({}^3P_1):\quad&
\varphi_{1^{++}}^{\para}(x)
&&=+\varphi_{1^{++}}^{\para}(1-x),\\
&
\varphi_{1^{++}}^{\perpdir}(x)
&&=-\varphi_{1^{++}}^{\perpdir}(1-x),
\end{alignedat}
\label{eq:sym_3P1}
\end{equation}
\begin{equation}
\begin{alignedat}{2}
1^{+-}\ ({}^1P_1):\quad&
\varphi_{1^{+-}}^{\para}(x)
&&=-\varphi_{1^{+-}}^{\para}(1-x),\\
&
\varphi_{1^{+-}}^{\perpdir}(x)
&&=+\varphi_{1^{+-}}^{\perpdir}(1-x).
\end{alignedat}
\label{eq:sym_1P1}
\end{equation}
In this way, in the flavour-symmetric limit, the longitudinal DA is symmetric for the $1^{++}$ channel and antisymmetric for the $1^{+-}$ channel, whereas the transverse DA displays the opposite pattern. A formal proof is presented in Appendix\,\ref{app:Charge}. In mixed-flavour systems, such as strange AV mesons, these relations are no longer exact, and the distributions deviate from their ideal symmetry/antisymmetry patterns. Such deviations therefore provide a useful measure of the impact of $SU_F(3)$ flavour breaking.

The normalisation requires comment. It is fixed consistently with the
decay constants appearing in
Eqs.~\eqref{eq:DA_Vpar}--\eqref{eq:DA_Vperp}: for symmetric DAs,
\begin{equation}
\int_0^1 \dd x\,\varphi(x)=1\,.
\label{eq:sym_norm}
\end{equation}
For antisymmetric DAs, the zeroth moment vanishes identically:
\begin{equation}
\int_0^1 \dd x\,\varphi(x)=0\,.
\label{eq:antisym_zero}
\end{equation}
The overall scale and sign are then fixed by the standard weighted
normalisation\,\cite{Yang:2007zt}
\begin{equation}
\int_0^1 \dd x\, (1-2x)\,\varphi(x)=1\,,
\label{eq:antisym_norm}
\end{equation}
with which each sign-changing DA is positive on $x<1/2$.

We can distinguish the light-front projection from its normalisation as follows. Let us denote the right-hand sides of Eqs.\,\eqref{eq:DA_Vpar}-\eqref{eq:DA_Vperp} by $\mathcal P^{\para,\perpdir}(x)$. These represent the unnormalised projections, with the prefactors fixed by the chosen normalisation of $\varphi$. For a symmetric distribution, Eq.~\eqref{eq:sym_norm} yields:
\begin{equation}
f_{\text{AV}}=\int_0^1\dd x\,\mathcal P_{\text{AV}}(x),
\qquad
\varphi(x)=\mathcal P_{\text{AV}}(x)/f_{\text{AV}} \,,
\label{eq:normsym}
\end{equation}
where $f$ denotes the physical decay constant. For a sign-changing distribution, this normalization procedure is not applicable. In the flavour-symmetric limit, charge conjugation requires $\int_0^1\dd x,\mathcal P(x)=0$, so the corresponding coupling vanishes -- whence the entry $f_{b_1}=0$ in Table~\ref{TabMeson}. This cannot be used to fix the normalization. Instead, we employ Eq.~\eqref{eq:antisym_norm} and define:
\begin{equation}
\tilde f_{\text{AV}}=\int_0^1\dd x\,(1-2x) \mathcal P(x),
\qquad
\varphi(x)=\mathcal P(x)/\tilde f_{\text{AV}} \,,
\label{eq:normantisym}
\end{equation}
where $\tilde f_{\text{AV}}$ has the dimensions of a decay constant but cannot be identified as a physical observable.
The physical coupling is then recovered from the normalized distribution according to $f_{\text{AV}}=\tilde f_{\text{AV}}\int_0^1\dd x,\varphi_{\text{AV}}(x)$.
Thus, $f_{\text{AV}}=\tilde f_{\text{AV}}$ only when the integral of the normalized distribution equals unity. In the flavour-symmetric limit, the integral vanishes by charge conjugation, consistent with the vanishing of the corresponding physical coupling. In the strange channels, however, the integral is nonzero, and the physical coupling can be recovered from Eq.~\eqref{eq:normantisym}. For $K_1^{+-,\para}$, for example:
\begin{eqnarray}
    \tilde{f}_{K_1^{+-}}=0.066\,\text{GeV}\,,\,\int_0^1 dx\, \varphi_{K_1^{+-}}(x)&=&0.289 \\
    \Rightarrow f_{K_1^{+-}}&=&0.019\,\text{GeV}\,.\nonumber
\end{eqnarray}
Decay constants are listed in Table~\ref{TabMeson}, and every entry in the $f_{\rm calc.}$ column corresponds to a physical decay constant. Note that Eq.~\eqref{eq:antisym_norm} should be understood as a normalisation convention: it fixes the shape and overall sign of each sign-changing distribution but not its physical magnitude, which resides in the product $\tilde f\,\varphi(x)$. Comparisons between sign-changing channels made below are consequently comparisons of their normalized shapes, under the same weighted normalization condition.

Having established the definitions and normalisation conventions, the next section describes the computation of the required BSWFs within an improved, symmetry-preserving CSMs treatment.

\section{Gap and Bethe--Salpeter equations}
\label{sec:gap_bse}

In order to compute the DAs defined in
Eqs.~\eqref{eq:DA_Vpar}-\eqref{eq:DA_Vperp}, one requires the
dressed-quark propagators and the Bethe--Salpeter amplitudes of the
relevant vector and AV mesons. The quark propagators are
obtained from the QCD gap equation \cite{Roberts:1994dr}. They then
enter the homogeneous Bethe--Salpeter equation, whose solution yields
the bound-state amplitudes used in the light-front projections. The two
equations cannot be specified independently: the axial-vector
Ward--Green--Takahashi identity imposes a close connection between the
gap-equation kernel and the Bethe--Salpeter kernel
\cite{Munczek:1994zz,Bender:1996bb}. In contemporary continuum
Schwinger-function methods, this connection is implemented
systematically. Once an effective charge and a dressed gluon--quark
vertex are specified in the gap equation, the associated
symmetry-consistent Bethe--Salpeter kernel can be constructed
\cite{Qin:2020jig,Xu:2022kng,Xu:2025cyj,Xiao:2025cqz,Yao:2025xjx}.

Analyses of QCD's gauge sector,\,\cite{Binosi:2016nme,Rodriguez-Quintero:2018wma,Cui:2019dwv},
have delivered a process-independent (PI) effective charge, the QCD
analogue of the Gell-Mann--Low charge in QED\,\cite{Gell-Mann:1954yli}. Its recent developments and applications are
discussed in Refs.~\cite{Ferreira:2025anh,Brodsky:2024zev,Deur:2023dzc}, including its connection with process-dependent effective charges. Herein, we employ a bottom-up representation of the PI charge,
constrained by hadron observables\,\cite{Binosi:2014aea}, so that the
quark gap equation can be written as follows\,\cite{Chang:2021vvx}:

\begin{subequations}
\label{EqGap}
\begin{align}
S^{-1}(k) &= i\gamma\cdot k + m + \Sigma(k) \,,
\\
\Sigma(k) &=
\int\frac{\dd^4 q}{(2\pi)^4}
\,4 \pi {\mathpzc A}_{\mu\nu}(l)
\,\gamma_\mu\frac{\lambda^{a}}{2} S(q)
\Gamma_\nu(q,k)\frac{\lambda^{a}}{2}\,.
\end{align}
\end{subequations}
Herein, all dependence on the renormalisation scale $\zeta$ is
suppressed. Our renormalisation procedure is described, for example, in
Ref.~\cite{Chang:2008ec}; it is straightforward to implement and
preserves the relevant symmetries. In Eq.~\eqref{EqGap}, $l=k-q$ and
$m$ is the quark current mass. The matrices
$\{\lambda^a/2\,|\,a=1,\ldots,8\}$ are the generators of SU$(3)$ colour
in the fundamental representation, ${\mathpzc A}_{\mu\nu}$ is the
effective vector-boson exchange interaction, and $\Gamma_\nu$ is the
relevant dressed gluon--quark vertex. The solution of Eq.~\eqref{EqGap} is written as:
\begin{equation}
S(k)=\frac{1}{i\gamma\cdot k\,A(k^2)+B(k^2)}.
\label{eq:quark_prop_AB}
\end{equation}

Our approximation to the PI charge is introduced by writing\,\cite{Qin:2011dd}:
\begin{subequations}
\label{Amunu}
\begin{align}
{\mathpzc A}_{\mu\nu}(l)
&=
T_{\mu\nu}(l)\,\tilde{\mathpzc A}(y=l^2),
\\
\tilde{\mathpzc A}(y)
&=
2\pi \frac{D}{\omega^4} {\rm e}^{-y/\omega^2}
+
\frac{2\pi \gamma_m \mathcal{F}(y)}
{\ln\!\left[\tau+\left(1+y/\Lambda_{\rm QCD}^2\right)^2\right]} \,.
\end{align}
\end{subequations}
The quantities entering the ultraviolet contribution are given by:
\begin{equation}
\begin{aligned}
\mathcal F(y)
&=
\frac{1-\exp(-y/\Lambda_{\mathpzc I}^{2})}{y},
\quad
\gamma_m=\frac{12}{25}\,,
\\
\Lambda_{\rm QCD}
&=0.234\,{\rm GeV},
\quad
\tau={\rm e}^2-1,
\quad
\Lambda_{\mathpzc I}=1\,{\rm GeV}.\,
\end{aligned}
\label{eq:Amunu_parameters}
\end{equation}
This interaction is not a pointwise approximation to the QCD running
coupling on the far infrared domain. Rather, it is designed to reproduce
global properties of the interaction on the momentum domain relevant to
hadron bound-state calculations. The tensor structure
$T_{\mu\nu}(l)=\delta_{\mu\nu}-l_\mu l_\nu/l^2$ is the standard
transverse projector associated with Landau gauge, which we use because
it is a fixed point of the renormalisation group.

The quark-gluon vertex is written as ($l=k-q$):
\begin{equation}
\Gamma_\nu(q,k)
=
\gamma_\nu
+
\eta\,\kappa(l^2)\,l_\alpha\sigma_{\alpha\nu}\,,
\label{eq:acm_vertex}
\end{equation}
where the momentum-dependent factor associated with the anomalous chromomagnetic moment (ACM) is:
\begin{equation}
\kappa(l^2)=\frac{1}{\omega}\exp(-l^2/\omega^2)\,.
\label{eq:kappa_acm}
\end{equation}
A possible overall multiplicative factor is implicitly absorbed into
$\tilde{\mathpzc A}$. Moreover, with $\eta=0$, Eq.~\eqref{eq:acm_vertex}, one recovers the RL truncation. The latter is known to be
reliable for ground-state hadrons with little rest-frame orbital angular
momentum, but it must be improved for channels in which spin-orbit
effects are important, such as excited states and AV mesons
\cite{Fischer:2009jm,Chang:2009zb,Qin:2011xq,Chang:2011ei}. For $\eta>0$, the ACM term provides a valuable extension beyond RL, incorporating spin-dependent interactions associated with EHM\,\cite{Chang:2010hb}. Its inclusion has been shown to improve the description of meson spectra and leptonic decays involving light and strange quarks\,\cite{Xu:2022kng}.

Following Ref.~\cite{Xu:2025cyj,Xiao:2025cqz}, we employ:
\begin{equation}
\omega=0.8\,{\rm GeV},
\qquad
D=(0.68\,{\rm GeV})^2,
\qquad
\eta=1.1,
\label{eq:model_parameters}
\end{equation}
together with the renormalisation-point-invariant current-quark masses:
\begin{equation}
\hat m_u=\hat m_d=4.07\,{\rm MeV},
\qquad
\hat m_s=110\,{\rm MeV}.
\label{eq:rgi_masses}
\end{equation}
These values correspond to one-loop current masses at
$\zeta=2\,{\rm GeV}$ of:
\begin{equation}
m_u=2.82\,{\rm MeV},
\qquad
m_s=76.2\,{\rm MeV}.
\label{eq:zeta2_masses}
\end{equation}

For a meson with total momentum $P$, the BSA $\Gamma_\nu(k,P)$ satisfies
\cite{Maris:1997tm,Maris:2005tt,Krassnigg:2008bob}
\begin{equation}
\Gamma_\nu(k,P)
=
\lambda(P^2)
\int^\Lambda \frac{\dd^4q}{(2\pi)^4}\,
K^{(2)}(k,q;P)\,
\chi_\nu(q,P),
\label{eq:BSE_general}
\end{equation}
where $\lambda(P^2)$ is the eigenvalue and $\int^\Lambda$ denotes a
translationally invariant regularisation. The kernel is not independent
of the gap equation: symmetry consistency requires that it be built from
the same dressed gluon--quark vertex. We use the closed-form construction
of Refs.~\cite{Qin:2020jig,Xu:2022kng}, which yields a kernel consistent
with any reasonable $\Gamma_\nu$, whether or not its diagrammatic content
is known.

Writing $\mathcal G_{\mu\nu}(l)=4\pi{\mathpzc A}_{\mu\nu}(l)$ and
denoting the ACM term of Eq.~\eqref{eq:acm_vertex} by
$\tau_\nu(l)=\eta\,\kappa(l^2)l_\alpha\sigma_{\alpha\nu}$, the
two-particle-irreducible quark+antiquark scattering kernel is
\begin{equation}
\begin{aligned}
K^{(2)} =\;
&-\mathcal G_{\mu\nu}(l)\,\gamma_\mu\otimes\gamma_\nu
-\mathcal G_{\mu\nu}(l)\,\gamma_\mu\otimes\tau_\nu(l)
\\
&+\mathcal G_{\mu\nu}(l)\,\tau_\nu(l)\otimes\gamma_\mu
+K_{\rm ad},
\end{aligned}
\label{eq:K2}
\end{equation}
in which the first term is the ladder kernel, the next two are the
minimal ACM insertions on either leg, and
\begin{equation}
\begin{aligned}
K_{\rm ad}=\;
&\big[\mathbf 1\otimes_+\mathbf 1\big]f^{(+)}_{p0}
+\big[-\mathcal G_{\mu\nu}(l)\gamma_\mu\otimes_+\gamma_\nu\big]
 f^{(-)}_{p1}
\\
&+\big[\mathbf 1\otimes_-\mathbf 1\big]f^{(+)}_{n0}
+\big[-\mathcal G_{\mu\nu}(l)\sigma_{l\mu}\otimes_-\sigma_{l\nu}
 \big]f^{(+)}_{n1},
\end{aligned}
\label{eq:Kad}
\end{equation}
with $\otimes_\pm:=\tfrac12(\otimes\pm\gamma_5\otimes\gamma_5)$ and
$\sigma_{l\mu}=\sigma_{\rho\mu}l_\rho$. The scalar functions
$f^{(\pm)}_{pj}$ and $f^{(+)}_{nj}$, $j=0,1$, are not chosen but
determined by the vector and axial--vector Ward--Green--Takahashi
identities. Defining $q_\pm=q\pm P/2$ and
$\int_{dq}:=\int^\Lambda\dd^4q/(2\pi)^4$, they solve
\begin{subequations}
\label{eq:Kconstraints}
\begin{align}
&\int_{dq}\mathcal G_{\mu\nu}(l)\gamma_\mu s^g_A(q_+)\tau_\nu(l)
\nonumber\\
&\quad=\int_{dq}\Big[s^g_B(q_+)f^{(+)}_{p0}
 +\mathcal G_{\mu\nu}(l)\gamma_\mu s^h_B(q_-)\gamma_\nu
 f^{(-)}_{p1}\Big],
\\
&\int_{dq}\mathcal G_{\mu\nu}(l)\gamma_\mu s^g_B(q_+)\tau_\nu(l)
\nonumber\\
&\quad=\int_{dq}\Big[s^g_A(q_+)f^{(+)}_{n0}
 -\mathcal G_{\mu\nu}(l)\sigma_{l\mu}s^g_A(q_+)\sigma_{l\nu}
 f^{(+)}_{n1}\Big].
\end{align}
\end{subequations}
Here $g,h$ label the valence flavours and
$S_g(k)=:s^g_A(k)+s^g_B(k)$, with
$\{s^g_A,\gamma_5\}=0=[s^g_B,\gamma_5]$. The solutions depend on
$P^2$; they are obtained on a $P^2$ grid and interpolated. For $u,d$
channels in the isospin-symmetric limit, and separately for $s\bar s$,
Eqs.~\eqref{eq:Kconstraints} yield four real-valued functions each;
six are required for $u\bar s$ and kindred channels, because when
$m_g\ne m_h$ the identities satisfied by $\Sigma^g_B(k_+)$ and
$\Sigma^h_B(k_-)$ are no longer related by charge
conjugation\,\cite{Xu:2022kng}.

Because the $f$'s are fixed by resolving the WGT identities, those
identities are preserved by construction for any $\eta$; the
Gell-Mann--Oakes--Renner and Goldberger--Treiman relations
follow~\cite{Qin:2020jig,Xu:2022kng}. Setting $\eta=0$ gives
$\tau_\nu\equiv0$, so that the left-hand sides of
Eqs.~\eqref{eq:Kconstraints} vanish; the residual homogeneous
equations are solved by $f\equiv0$, whence $K_{\rm ad}=0$ and
Eq.~\eqref{eq:K2} collapses to the ladder kernel
$-\mathcal G_{\mu\nu}(l)\gamma_\mu\otimes\gamma_\nu$. The parameter
$\eta$ therefore interpolates continuously between RL truncation and
the EHM-improved kernel used here.

The physical mass is determined from
\begin{equation}
\lambda(P^2=-m_H^2)=1\,.
\label{eq:eigen_mass}
\end{equation}
The BSA is then canonically normalised\,\cite{Nakanishi:1969ph}, and the
resulting BSWF is used in the decay-constant and DA projections.
Consequently, the spectrum, decay constants, and Mellin moments follow
from the same self-consistent set of dressed-quark propagators and BSAs.

The masses and decay constants relevant to the present study are collected in Table~\ref{TabMeson}. It provides the meson properties entering the kinematics of the DA projections. It also demonstrates that the Bethe-Salpeter amplitudes used to compute the DA moments are associated with a phenomenologically sound description of the relevant vector and AV channels. The pseudoscalar entries are included as benchmarks. Where available, empirical values are taken from the PDG~\cite{ParticleDataGroup:2024cfk}.

The AV channels provide the sharpest test of the interaction kernel, since these are precisely the states for which the RL truncation fails badly: it underestimates $m_{a_1}$ and $m_{b_1}$ by $27\%$ and $33\%$, respectively~\cite{Ferreira:2026gbe}. With the vertex of Eq.~\eqref{eq:acm_vertex}, these discrepancies are reduced to $2.4\%$ and $4.9\%$, while the strange AV masses lie within about $5\%$ of the unmixed basis-state values quoted in Table~\ref{TabMeson}. This level of agreement with the AV spectrum provides important support for the Bethe-Salpeter kernel and, in turn, confidence in the BSWFs from which the DA moments are computed below.

With the required BSWFs at hand, and with the phenomenological support provided by the resulting mass spectrum and decay constants, we now turn to the calculation of the DA Mellin moments.

\begin{table}[t]
\centering
\caption{Masses and decay constants, in GeV, for the mesons relevant to
the present DA calculation. The calculated values are obtained using
the interaction and vertex specified in Eqs.~\eqref{Amunu} and
\eqref{eq:acm_vertex}, with parameters listed in
Eqs.~\eqref{eq:model_parameters}--\eqref{eq:rgi_masses}. Empirical
values are quoted from the PDG~\cite{ParticleDataGroup:2024cfk} where
available.
$^{\rm a}$For the strange axial--vector channels the entries in the
third column are not the physical $K_1(1400)$ and $K_1(1270)$ masses.
As explained in Sec.~\ref{sec:intro}, $K_1^{++}$ and $K_1^{+-}$ are the
unmixed partners of the flavour-symmetric $1^{++}$ and $1^{+-}$
channels, whereas the physical states are near-maximal admixtures of
them. We therefore quote instead the masses of the unmixed $K_{1A}$,
$K_{1B}$ basis states as extracted in Ref.~\cite{Ferreira:2026gbe} from
the physical spectrum using the mixing angle
$\theta_K=(40\pm5)^\circ$~\cite{Liu:2024lph}.}
\label{TabMeson}
\renewcommand{\arraystretch}{1.10}
\setlength{\tabcolsep}{5.0pt}
\begin{tabular}{lcccc}
\toprule
Meson & $m_{\mathrm{calc.}}$ & $m_{\mathrm{exp.}}$
& $f_{\mathrm{calc.}}$ & $f_{\mathrm{exp.}}$ \\
\midrule
$\pi$        & 0.14  & 0.14      & 0.098 & 0.092    \\
$\rho$       & 0.77  & 0.77      & 0.159 & 0.153(1) \\
$K$          & 0.494 & 0.494     & 0.110 & 0.110(1) \\
$K^\ast$     & 0.868 & 0.892     & 0.178 & 0.159(1) \\
$a_1$        & 1.20  & 1.23(4)   & 0.136 & --       \\
$b_1$        & 1.17  & 1.230(3)  & 0     & --       \\
$K_1^{++}$   & 1.304 & 1.343(17)$^{\rm a}$ & 0.147 & --       \\
$K_1^{+-}$   & 1.25  & 1.317(17)$^{\rm a}$ & 0.019 & --       \\
\bottomrule
\end{tabular}
\end{table}

\section{Mellin moments and reconstructed DAs}
\label{sec:reconstruct}

\subsection{Mellin moments}

The light-front projections defined in Eqs.~\eqref{eq:DA_Vpar} and
\eqref{eq:DA_Vperp} yield the Mellin moments, which are the primary
quantities calculated here. Tables~\ref{tab:DA-moments-light} and
\ref{tab:DA-moments-strange} collect the results for the
flavour-symmetric ($\rho$, $a_1$, $b_1$) and strange ($K^\ast$,
$K_1^{++}$, $K_1^{+-}$) channels, respectively. All results refer to
the hadron scale, $\zeta_H$, associated with the gap and Bethe--Salpeter framework\,\cite{Cui:2020tdf,Cui:2020dlm}
of Sec.~\ref{sec:gap_bse}, and also defined in connection with the process-independent effective charges\,\cite{Cui:2019dwv,Binosi:2016nme} (see, e.g., Ref.\,\cite{Deur:2023dzc} for a review of the theory of effective charges).

\begin{table}[htbp]
\centering
\caption{Longitudinal and transverse Mellin moments
($m=0,\ldots,6$) for the flavour-symmetric channels
$\rho$, $a_1$, and $b_1$.}
\label{tab:DA-moments-light}
\small
\renewcommand{\arraystretch}{1.10}
\setlength{\tabcolsep}{5.0pt}
\begin{tabular}{ccccccc}
\toprule
$m$
& \multicolumn{2}{c}{$\rho$}
& \multicolumn{2}{c}{$a_1$}
& \multicolumn{2}{c}{$b_1$}
\\
\cmidrule(lr){2-3}
\cmidrule(lr){4-5}
\cmidrule(lr){6-7}
&
$\langle x^m\rangle^{\para}$
& $\langle x^m\rangle^{\perpdir}$
& $\langle x^m\rangle^{\para}$
& $\langle x^m\rangle^{\perpdir}$
& $\langle x^m\rangle^{\para}$
& $\langle x^m\rangle^{\perpdir}$
\\
\midrule
0 & 1     & 1     & 1     & 0      & 0      & 1     \\
1 & 0.5   & 0.5   & 0.5   & -0.5   & -0.5   & 0.5   \\
2 & 0.295 & 0.296 & 0.282 & -0.5   & -0.5   & 0.284 \\
3 & 0.193 & 0.194 & 0.174 & -0.423 & -0.424 & 0.177 \\
4 & 0.136 & 0.135 & 0.114 & -0.345 & -0.347 & 0.116 \\
5 & 0.100 & 0.099 & 0.079 & -0.281 & -0.285 & 0.084 \\
6 & -- & --    & 0.056 & -0.231 & -0.237 & 0.065 \\
\bottomrule
\end{tabular}
\end{table}

\begin{table}[htbp]
\centering
\caption{Longitudinal and transverse Mellin moments
($m=0,\ldots,6$) for the strange channels
$K^\ast$, $K_1^{++}$, and $K_1^{+-}$.}
\label{tab:DA-moments-strange}
\small
\renewcommand{\arraystretch}{1.10}
\setlength{\tabcolsep}{5.0pt}
\begin{tabular}{ccccccc}
\toprule
$m$
& \multicolumn{2}{c}{$K^\ast$}
& \multicolumn{2}{c}{$K_1^{++}$}
& \multicolumn{2}{c}{$K_1^{+-}$}
\\
\cmidrule(lr){2-3}
\cmidrule(lr){4-5}
\cmidrule(lr){6-7}
&
$\langle x^m\rangle^{\para}$
& $\langle x^m\rangle^{\perpdir}$
& $\langle x^m\rangle^{\para}$
& $\langle x^m\rangle^{\perpdir}$
& $\langle x^m\rangle^{\para}$
& $\langle x^m\rangle^{\perpdir}$
\\
\midrule
0 & 1     & 1     & 1     & 0.177  & 0.289  & 1      \\
1 & 0.524 & 0.529 & 0.532 & -0.410 & -0.356 & 0.513  \\
2 & 0.319 & 0.323 & 0.316 & -0.475 & -0.442 & 0.297  \\
3 & 0.212 & 0.216 & 0.202 & -0.430 & -0.409 & 0.186  \\
4 & 0.151 & 0.148 & 0.136 & -0.367 & -0.353 & 0.124  \\
5 & 0.115 & 0.121 & 0.096 & -0.309 & -0.299 & 0.0846 \\
6 & -- & --    & 0.070 & -0.260 & -0.251 & 0.0613 \\
\bottomrule
\end{tabular}
\end{table}

For vector mesons, the longitudinal and transverse moments differ by
less than one per cent for the $\rho$ and by about two per cent for the
$K^\ast$ up to $m=4$. This suppressed polarisation splitting is not
imposed by symmetry; it is a dynamical result. In the AV sector, charge
conjugation instead gives
$\langle x^0\rangle_{a_1}^{\perpdir}=0$ and
$\langle x^0\rangle_{b_1}^{\para}=0$, whereas flavour breaking produces
$\langle x^0\rangle_{K_1^{++}}^{\perpdir}=0.177$ and
$\langle x^0\rangle_{K_1^{+-}}^{\para}=0.289$. For the exactly
antisymmetric channels, the entries with $m=0,1,2$ are fixed by
Eqs.~\eqref{eq:sym_3P1}--\eqref{eq:antisym_norm}; the first independent
moment is $\langle x^3\rangle$.

For positive-definite channels, a useful measure of skewness is
\begin{equation}
\langle \xi\rangle =
\langle 1-2x\rangle =
1-2\langle x\rangle,
\label{eq:xi_moment}
\end{equation}
which vanishes for a DA symmetric about $x=1/2$. With $x$ assigned to
the heavier valence quark, a negative value indicates that it carries
the larger share of the light-front momentum. Table~\ref{tab:DA-moments-strange}
gives
\begin{equation}
\begin{gathered}
\langle \xi\rangle_{K^{*\para}}=-0.048,\qquad
\langle \xi\rangle_{K^{*\perpdir}}=-0.058,\\
\langle \xi\rangle_{K_1^{++\,\para}}=-0.064,\qquad
\langle \xi\rangle_{K_1^{+-\,\perpdir}}=-0.026.
\end{gathered}
\label{eq:xi_values}
\end{equation}
Measured by $\langle\xi\rangle$, flavour breaking is of comparable
overall size in the two sectors, but its channel-to-channel variation
is about four times larger for AV mesons.
The qualitative distinction is that SU(3) breaking lifts a
charge-conjugation-enforced zero in the AV sector; neither vector
projection is forbidden in the symmetric limit.

In the flavour-symmetric channels, charge conjugation fixes the
reflection parity of the DAs under $x\leftrightarrow1-x$
\cite{Yang:2007zt}. Applying this transformation to the definition of
the Mellin moments and using the corresponding normalisations gives,
for symmetric DAs,
\begin{equation}
\langle x^3\rangle
 =-\tfrac14+\tfrac32\langle x^2\rangle,
\qquad
\langle x^5\rangle
 =\tfrac12-\tfrac52\langle x^2\rangle
  +\tfrac52\langle x^4\rangle,
\label{eq:recurrence-sym}
\end{equation}
and, for antisymmetric DAs,
\begin{equation}
\langle x^4\rangle
 =\tfrac12+2\langle x^3\rangle,
\qquad
\langle x^6\rangle
 =\tfrac94+10\langle x^3\rangle
  -\tfrac{15}{2}\langle x^4\rangle
  +3\langle x^5\rangle.
\label{eq:recurrence-anti}
\end{equation}
Here, the antisymmetric relations also use the weighted normalisation
in Eq.~\eqref{eq:antisym_norm}. The lower-order identities, involving
moments through $m=4$, are satisfied to better than $0.6\%$. The
higher-order identities involving $m=5,6$ are satisfied to better than
$2.5\%$, except for the fifth-moment identity of $b_1^{\perpdir}$,
whose residual is $4.8\%$. As independent checks, all positive-definite
DAs satisfy the Cauchy--Schwarz inequality
\begin{equation}
\langle x^{m+1}\rangle^2
 \leq \langle x^m\rangle\langle x^{m+2}\rangle,
\end{equation}
and all four sign-changing DAs reproduce
$\langle x^0\rangle-2\langle x^1\rangle=1$ to within $0.3\%$.

\subsection{DA reconstruction}

We reconstruct the DAs by fitting analytic forms to the moments in
Tables~\ref{tab:DA-moments-light} and~\ref{tab:DA-moments-strange}.
The parameters are obtained from fits to all available moments. The
agreement is best at low orders: the reconstructed moments reproduce
the direct Bethe--Salpeter results to better than one per cent through
$m=4$, except for $K^{*\perpdir}$, for which this accuracy is maintained
through $m=3$. Deviations increase at higher orders, which are more
sensitive to numerical uncertainty and endpoint behaviour.

Two qualitatively different DA profiles occur. Some channels are smooth,
positive-definite, and single-peaked; others possess a sign-changing
structure, enforced by charge conjugation in the flavour-symmetric
channels and inherited from that symmetry limit in the strange
channels. We therefore use two complementary parametrisations.

For sign-changing channels, we use the Gegenbauer representation, with
$\xi=1-2x$ as in Eq.~\eqref{eq:xi_moment}:
\begin{equation}
\begin{aligned}
\varphi(x)
&=
6x(1-x)
\Big[
a_0+a_1 C_1^{3/2}(\xi)
\\
&\quad
+a_2 C_2^{3/2}(\xi)
+a_3 C_3^{3/2}(\xi)
\Big].
\end{aligned}
\label{eq:gegen}
\end{equation}
The $C_n^{3/2}$ polynomials preserve the standard QCD weight
$6x(1-x)$ and implement reflection symmetry through their parity
under $\xi\to-\xi$~\cite{Lepage:1980fj}. We use Eq.~\eqref{eq:gegen}
for $a_1^{\perpdir}$, $b_1^{\para}$, $K_1^{++\,\perpdir}$, and
$K_1^{+-\,\para}$. Charge conjugation enforces the midpoint node in
the first two channels; in the strange channels, even Gegenbauer
components encode the loss of exact antisymmetry. Higher terms are not
stably constrained by the available moments.

For smooth positive-definite channels, we use an exponential
moment-reconstruction form employed in continuum Bethe--Salpeter studies
of meson DAs~\cite{Ding:2015rkn,Serna:2020txe,Raya:2024ejx,Raya:2026gwg}. For
$x\in[0,1]$, it reads
\begin{equation}
\begin{aligned}
\varphi(x)
&=\mathcal N_P\,x(1-x)
\\
&\quad\times\exp\left[
\frac{x(1-x)}{\rho_P^2}
+\gamma_P(1-2x)
\right].
\end{aligned}
\label{eq:exp-ansatz}
\end{equation}
The factor $x(1-x)$ gives the endpoint suppression. The parameters are
determined by fitting the calculated Mellin moments: $\mathcal N_P$
controls the overall normalisation, $\rho_P$ the width, and $\gamma_P$
the skewness. For flavour-symmetric channels, $\gamma_P=0$. Note that $\exp[x(1-x)/\rho_P^2]$ enhances the central region for any
real $\rho_P$, so Eq.~\eqref{eq:exp-ansatz} cannot describe
distributions broader than $\varphi_{\rm asy}$. This is adequate here
because every channel is found to satisfy $\langle\xi^2\rangle<1/5$,
a property established directly from the moments and independent of
the fit. Because the
zeroth moment is included in the fit, the reconstructed DAs reproduce
unit normalisation within numerical accuracy. We use this form for
$\rho^{\para}$, $\rho^{\perpdir}$, $K^{*\para}$,
$K^{*\perpdir}$, $a_1^{\para}$, $b_1^{\perpdir}$,
$K_1^{++\,\para}$, and $K_1^{+-\,\perpdir}$.

Table~\ref{tab:exp-fit-params} lists the fitted parameters. In the
strange channels, $\gamma_P\ne0$ encodes the asymmetry reflected by the
nonzero first moments. For the sign-changing DAs,
Eq.~\eqref{eq:antisym_norm} fixes $a_1=5/3$, while orthogonality gives
$a_0=\langle x^0\rangle$; $a_2$ and $a_3$ are then determined from the
higher moments. The even coefficients $a_0$ and $a_2$ vanish in the
flavour-symmetric channels and become nonzero when flavour symmetry is
broken. Table~\ref{tab:gegen-coeffs} lists the resulting coefficients.

\begin{table}[htbp]
\centering
\caption{Parameters of the compact exponential representation,
Eq.~\eqref{eq:exp-ansatz}, for the smooth positive-definite channels.}
\label{tab:exp-fit-params}
\small
\setlength{\tabcolsep}{10pt}
\renewcommand{\arraystretch}{1.15}
\begin{tabular}{lccc}
\toprule
Channel & $\mathcal N_P$ & $\rho_P$ & $\gamma_P$ \\
\midrule
$\rho^{\parallel}$ & 4.408 & 0.812 & 0 \\
$\rho^{\perp}$ & 4.315 & 0.783 & 0 \\
$K^{*\parallel}$ & 3.949 & 0.704 & -0.277 \\
$K^{*\perp}$ & 3.835 & 0.680 & -0.317 \\
$a_1^{\parallel}$ & 1.144 & 0.356 & 0 \\
$b_1^{\perp}$ & 1.382 & 0.377 & 0 \\
$K_1^{++\,\parallel}$ & 0.963 & 0.341 & -0.545 \\
$K_1^{+-\,\perp}$ & 1.091 & 0.351 & -0.219 \\
\bottomrule
\end{tabular}
\end{table}

\begin{table}[htbp]
\centering
\caption{Gegenbauer coefficients in Eq.~\eqref{eq:gegen} for the
sign-changing channels. Orthogonality fixes
$a_0=\langle x^0\rangle$ and Eq.~\eqref{eq:antisym_norm} fixes
$a_1=5/3$; $a_2$ and $a_3$ follow from the higher moments. We use
$\xi=1-2x$.}
\label{tab:gegen-coeffs}
\small
\setlength{\tabcolsep}{6pt}
\renewcommand{\arraystretch}{1.15}
\begin{tabular}{l c c c c c}
\toprule
State & Pol. & $a_{0}$ & $a_{1}$ & $a_{2}$ & $a_{3}$ \\
\midrule
$a_{1}(1^{++})$     & $\perp$     & 0      & 1.667  & 0      & -0.252 \\
$b_{1}(1^{+-})$     & $\parallel$ & 0      & 1.667  & 0      & -0.184 \\
$K_{1}(1^{++})$     & $\perp$     & 0.177  & 1.667  & -0.303 & -0.158 \\
$K_{1}(1^{+-})$     & $\parallel$ & 0.289  & 1.667  & -0.345 & -0.274 \\
\bottomrule
\end{tabular}
\end{table}

\subsection{Pointwise distributions}

\begin{figure}[t]
\centering
\includegraphics[width=\columnwidth]{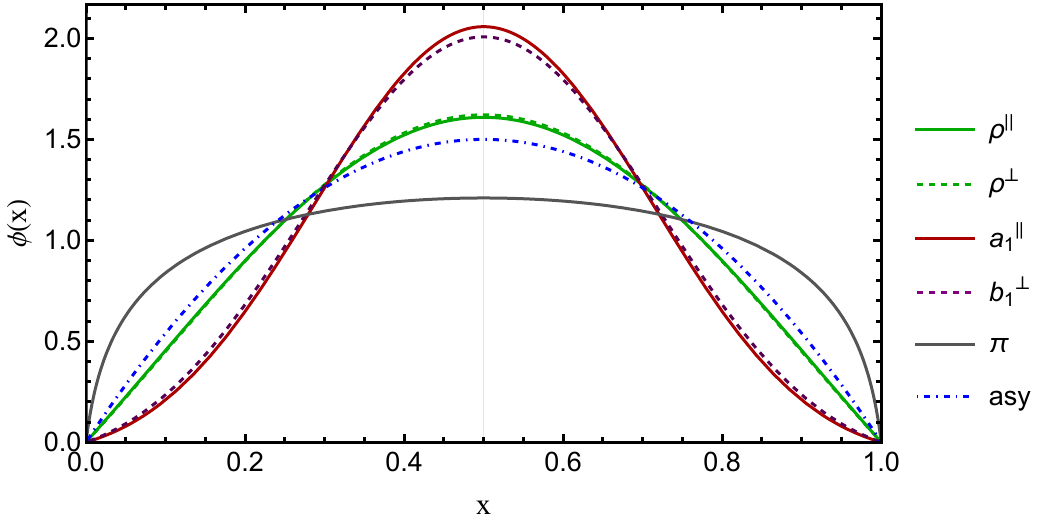}
\caption{Leading-twist DAs of flavour-symmetric positive-definite
channels, compared with the asymptotic distribution
$\varphi_{\rm asy}(x)=6x(1-x)$ and the pion DA.}
\label{fig:symm_positive}
\end{figure}

Figure~\ref{fig:symm_positive} shows the flavour-symmetric,
positive-definite DAs. The longitudinal and transverse $\rho$-meson
profiles are nearly identical. This near degeneracy is a dynamical
result, not a consequence of symmetry. Both lie between the broader
pion DA and the more strongly concentrated $a_1^\parallel$ and
$b_1^\perp$ profiles. Their second moments are
$(\langle\xi^2\rangle_\rho^\parallel,
\langle\xi^2\rangle_\rho^\perp)=(0.180,0.184)$, compared with
$0.128$ for $a_1^\parallel$ and $0.136$ for $b_1^\perp$. All are below
the asymptotic value $\langle\xi^2\rangle_{\rm asy}=1/5$, confirming
that these DAs are narrower than $\varphi_{\rm asy}(x)=6x(1-x)$ in the
second-moment sense. Evolution to larger resolving scales drives every
distribution toward the asymptotic limit, broadening those narrower
than $\varphi_{\rm asy}$ and narrowing those that are broader. This can
be done by applying ERBL evolution\,\footnote{For the flavour-symmetric sign-changing channels, ERBL
evolution preserves antisymmetry. With the weighted normalisation of
Eq.~\eqref{eq:antisym_norm}, $a_1=5/3$ remains fixed while the higher
odd components are progressively suppressed, so the DA approaches
$30x(1-x)(1-2x)$ at large scales.}, but this is beyond
the main purpose of this manuscript, namely, to express the
hadron-scale properties of the DA, especially the symmetry properties,
which remain unaltered under ERBL evolution.

\begin{figure}[t]
\centering
\includegraphics[width=\columnwidth]{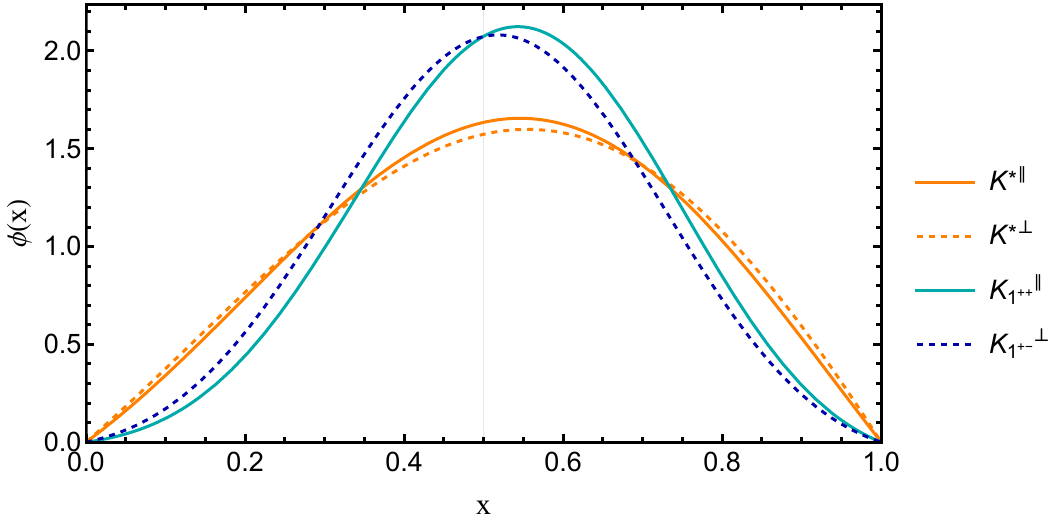}
\caption{Leading-twist DAs of the mixed-flavour positive-definite
channels: $K^{*\parallel}$, $K^{*\perp}$,
$K_1^{++\,\parallel}$, and $K_1^{+-\,\perp}$.}
\label{fig:mix_positive}
\end{figure}

In the mixed-flavour case, Fig.~\ref{fig:mix_positive}, the
$K^{*\parallel}$ and $K^{*\perp}$ DAs are mildly skewed but remain close
to one another. The strange axial channels show stronger deformation,
with $K_1^{++\,\parallel}$ exhibiting the largest first moment
$|\langle\xi\rangle|$ among the positive-definite channels. This contrast shows that the response to $SU_F(3)$ breaking is controlled
not by the current-mass splitting alone, but also by the
charge-conjugation structure of each channel: lifting a
symmetry-enforced zero produces a channel-dependent effect in AV
mesons, with no analogue in the vector sector.

\begin{figure}[t]
\centering
\includegraphics[width=\columnwidth]{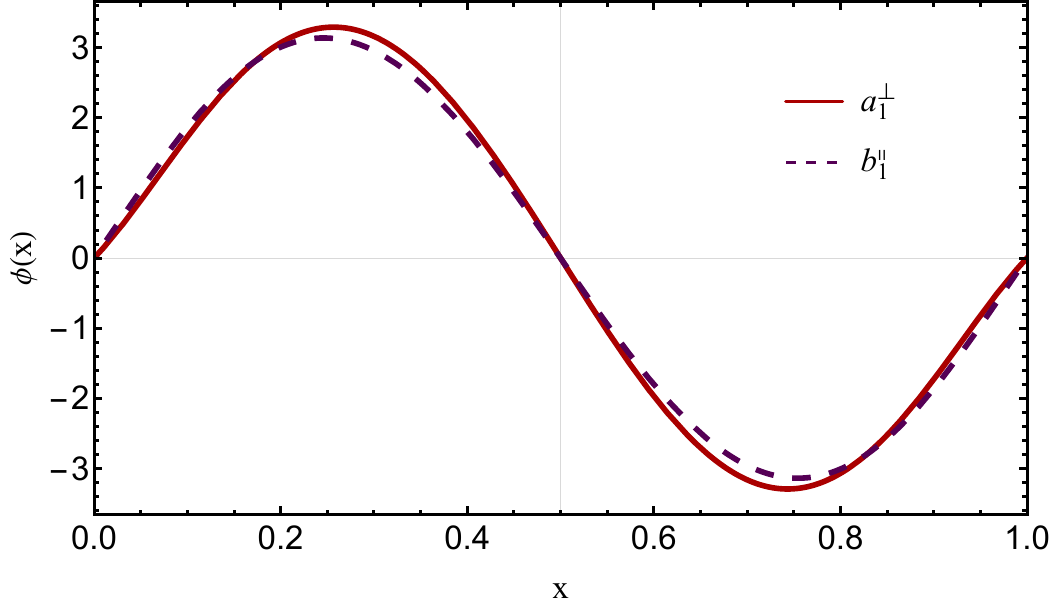}
\caption{Leading-twist DAs of the flavour-symmetric sign-changing
channels, $a_1^{\perpdir}$ and $b_1^{\para}$, displayed with the common
normalisation of Eq.~\eqref{eq:antisym_norm}.}
\label{fig:signchange_symm}
\end{figure}

\begin{figure}[t]
\centering
\includegraphics[width=\columnwidth]{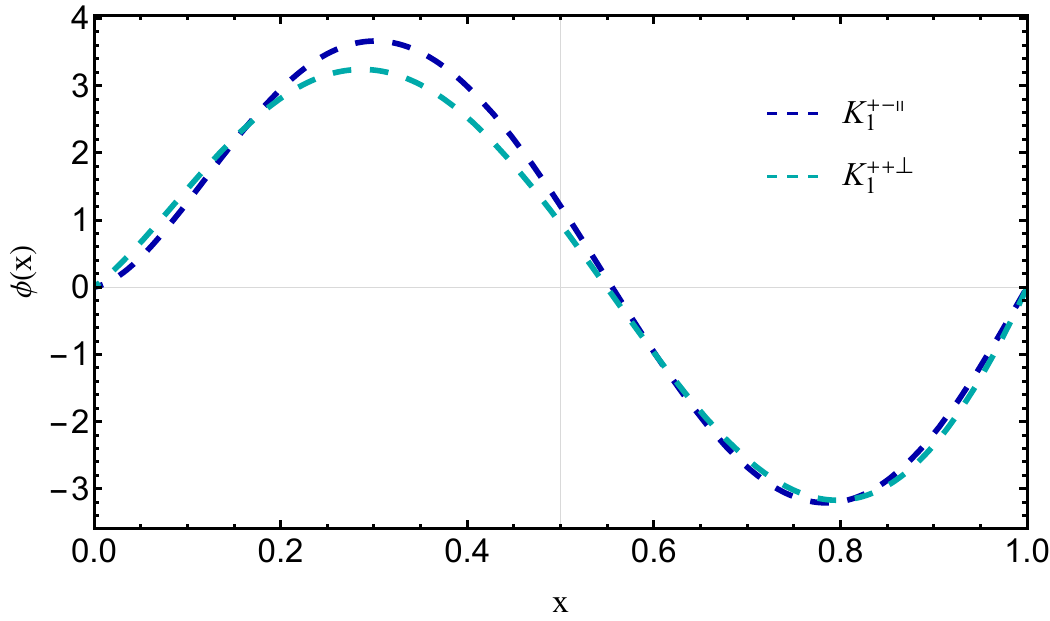}
\caption{Leading-twist DAs of the strange sign-changing channels,
$K_1^{++\,\perpdir}$ and $K_1^{+-\,\para}$. Both nodes lie at
$x\approx0.55$; the lobe asymmetry is larger for $K_1^{+-\,\para}$.}
\label{fig:signchange_strange}
\end{figure}

Figures~\ref{fig:signchange_symm} and~\ref{fig:signchange_strange}
show the sign-changing AV DAs. In the flavour-symmetric limit,
$a_1^{\perpdir}$ and $b_1^{\para}$ are antisymmetric and hence have
vanishing zeroth moments. Their common weighted normalisation fixes
$a_1=5/3$, while charge conjugation requires $a_0=a_2=0$. Their
different shapes are therefore encoded, at the order retained, in
$a_3/a_1=-0.151$ and $-0.110$, respectively. Both are close to the lowest antisymmetric Gegenbauer mode,
$\varphi_{\rm LO}^{\rm anti}(x)=30x(1-x)(1-2x)$, whose coefficient
follows from the weighted normalization in
Eq.~\eqref{eq:antisym_norm}. The modest $a_3$ contributions quantify
their departures from this symmetry-determined leading profile.

Flavour breaking shifts the balance of positive and negative support.
Both strange-channel nodes lie near $x=0.55$, but the imbalance between
the positive and negative lobes is more pronounced for
$K_1^{+-\,\para}$. With the weighted normalisation held fixed, the positive-lobe maximum
changes only from $3.26$ to $3.24$ in the $1^{++}$ channel, whereas it
increases from $3.14$ to $3.66$ in the $1^{+-}$ channel.  The zeroth moments show the same ordering:
$\langle x^0\rangle_{K_1^{+-}}^{\para}/
\langle x^0\rangle_{K_1^{++}}^{\perpdir}=1.63$.
A complementary signal is provided by the decay constant:
the longitudinal coupling forbidden for the $b_1$ in the
flavour-symmetric limit becomes
$f_{K_1^{+-}}=0.019\,\text{GeV}$ in the strange channel.

Comparison with the QCD sum-rule analysis of
Ref.~\cite{Yang:2007zt} shows agreement in the symmetry classification
and in the flavour-breaking first moments. The sum-rule system does not
determine the two $C$-violating zeroth moments independently and is
closed by assuming, in Eq.~(143) of that work, that the two ratios
$a_0/|a_1|$ are equal. For $K_1^{++\,\perpdir}$ and
$K_1^{+-\,\para}$, we instead obtain $0.106$ and $0.173$,
respectively, which are compatible with the quoted values $0.074(90)$
and $0.072(79)$ within their uncertainties.

The positive-definite AV channels differ more sharply. We obtain
$\langle\xi^2\rangle_{a_1}^{\para}=0.128$ and
$\langle\xi^2\rangle_{b_1}^{\perpdir}=0.136$, whereas the Gegenbauer
coefficients of Ref.~\cite{Yang:2007zt} correspond, via
$\langle\xi^2\rangle=1/5+(12/35)a_2$, to $0.193(7)$ and $0.210(65)$
at $\mu=1\,\text{GeV}$. Hence, the
$a_1^{\para}$ DA is substantially narrower in our calculation in the
second-moment sense. No equally sharp conclusion can be drawn for
$b_1^{\perpdir}$ because of the large sum-rule uncertainty. Since our
results refer to the hadron scale, a fully quantitative comparison
requires ERBL evolution to a common scale.

In the flavour-symmetric limit, charge conjugation assigns opposite
reflection parity to the longitudinal and transverse AV DAs. The
symmetric projections found here peak at $x=1/2$, whereas the
antisymmetric projections vanish there. This gives
$(\para,\perpdir)=(\text{peak},\text{node})$ for the $1^{++}$ channel,
with the assignment reversed for $1^{+-}$. Flavour breaking shifts
these features away from the midpoint. For $K_1^{++}$, the
$\para$ maximum and the $\perpdir$ node remain closely aligned at
$x\approx0.54$ and $0.55$, respectively. For $K_1^{+-}$, the
$\para$ node moves to $x\approx0.56$, farther than the $\perpdir$
maximum at $x\approx0.52$. The larger displacement of the
$K_1^{+-\,\para}$ node is accompanied by its larger zeroth moment and
even Gegenbauer components: three
features of the same profile, all pointing to a stronger
flavour-breaking response in the $1^{+-}$ channel.

\section{Conclusions and outlook}
\label{sec:concl}

We have computed the leading-twist DAs of vector and axial--vector
mesons using a nonperturbatively improved, symmetry-preserving
continuum-QCD framework, and reconstructed their pointwise
$x$-dependence from Mellin moments.

In the vector sector, the longitudinal and transverse DAs are nearly
degenerate: their moments agree to better than one per cent for the
$\rho$ up to $m=4$. No symmetry requires this, so the suppressed
polarisation dependence is a dynamical result of the calculated
Bethe--Salpeter amplitudes. Both $\rho$ DAs are slightly narrower than
the asymptotic distribution at the hadron scale, with
$(\langle\xi^2\rangle_\rho^{\para},
\langle\xi^2\rangle_\rho^{\perpdir})=(0.180,0.184)$, compared with
$\langle\xi^2\rangle_{\rm asy}=1/5$.

In the AV sector, the calculation reproduces the
charge-conjugation pattern that interchanges the symmetric and
antisymmetric projections between the $1^{++}$ and $1^{+-}$ channels.
Flavour breaking lifts the symmetry-enforced zeros unequally. The
resulting zeroth moments stand in the ratio $1.63$, and the positive
lobe of the $1^{+-}$ longitudinal projection is enhanced by $17$\%,
whereas its $1^{++}$ transverse counterpart changes by less than one
per cent. These two measures consistently show a stronger
flavour-breaking response in the $1^{+-}$ channel. A complementary
signal is provided by the longitudinal decay constant: the coupling
forbidden for the $b_1$ in the flavour-symmetric limit becomes
$f_{K_1^{+-}}^{\para}=0.019\,\text{GeV}$ in the strange channel. The
vector sector has no analogous effect because neither projection is
forbidden in the symmetric limit. What distinguishes the two sectors
is therefore the lifting of a symmetry-enforced zero, not simply the
size of the flavour-induced momentum asymmetry.

For the pointwise reconstruction, we use a compact exponential form for
the positive-definite DAs and a Gegenbauer expansion for the
sign-changing channels, whose odd and even components naturally encode
the node structure and its flavour-breaking distortion. Several
internal checks support the numerical consistency of the results. In
the flavour-symmetric channels, the calculated moments satisfy the
charge-conjugation recurrences of
Eqs.~\eqref{eq:recurrence-sym}--\eqref{eq:recurrence-anti} to better
than $0.6$\% through $m=4$. The positive-definite moments obey the
Cauchy--Schwarz inequalities, while all sign-changing channels recover
the weighted normalisation of Eq.~\eqref{eq:antisym_norm} to within
$0.3$\%.

Natural extensions are the implementation of physical
$K_1(1270)$--$K_1(1400)$ mixing and the evolution of the
hadron-scale DAs, so that the Gegenbauer coefficients may be quoted
at the scales conventionally employed in light-cone sum-rule
analyses. The distributions reported herein are direct inputs to
such analyses -- for instance, to $B\to K_1\gamma$, wherein the
photon polarisation in $b\to s\gamma$ is sensitive to the $K_1$
light-cone structure, and to $B\to a_1$ transitions; and a
systematic comparison of the present kernel with RL truncation would
isolate the influence of the dressed-quark anomalous chromomagnetic
moment on light-front momentum sharing.

A further step is the reconstruction of the associated light-front
wave functions, which would open the way to parton distributions,
transverse-momentum-dependent distributions and hard exclusive form
factors within a single framework. A matched comparison with RL
truncation, accompanied by variation of the interaction parameters,
would separate the effects of the dressed-quark anomalous
chromomagnetic moment from other model dependences and deliver a more
systematic uncertainty estimate. In the longer term, incorporating
the finite widths of the $a_1$ and $b_1$ would test the fidelity of
treating these broad resonances as stable Bethe--Salpeter bound
states.

\section*{Acknowledgments}
We are grateful to J.~Rodr\'iguez-Quintero for valuable discussions.
This work was supported by the Spanish Ministry of Science and
Innovation (MICINN) under Grant No.\ PID2022-140440NB-C22.
Z.Q.Y. acknowledges support from Helmholtz-Zentrum Dresden-Rossendorf
through the High Potential Programme.
P.C. acknowledges support from the Natural Science Foundation of
Anhui Province under Grant No.\ 2408085QA028.

\appendix

\section{Vector and axial--vector Bethe--Salpeter amplitudes}
\label{app:chebyshev}

We first specify the covariant decomposition of vector and AV
Bethe--Salpeter amplitudes (BSAs). For a given channel $X$, the amplitude is
expanded as
\begin{equation}
\Gamma_X(q,P) = \sum_i \tau_X^i(q,P)\,F_X^i(q^2,q\!\cdot\!P,P^2),
\label{eq:bsa-decomp}
\end{equation}
where $F_X^i$ are scalar dressing functions.

The vector-meson Bethe--Salpeter amplitude ($J^{PC}=1^{--}$) is written as
\begin{equation}
\Gamma_\mu^{1^-}(q;P)
=
\sum_{i=1}^{8} f_i^{1^-}(q^2,q\!\cdot\!P;P^2)\,
\tau^{i}_{1^-\mu}(q,P),
\end{equation}
with the transverse covariants
\begin{align}
\tau^{1}_{1^-\mu} &= i\gamma_\mu^T, \\
\tau^{2}_{1^-\mu} &= i\left[3q_\mu^T(\gamma^T\!\cdot\! q)-\gamma_\mu^T(q^T)^2\right], \\
\tau^{3}_{1^-\mu} &= i(P\!\cdot\! q)\,q_\mu^T\,\gamma\!\cdot\! P, \\
\tau^{4}_{1^-\mu} &= i\left[\gamma_\mu^T\,\gamma\!\cdot\! P\,(\gamma^T\!\cdot\! q)
      + q_\mu^T\,\gamma\!\cdot\! P\right], \\
\tau^{5}_{1^-\mu} &= q_\mu^T, \\
\tau^{6}_{1^-\mu} &= (P\!\cdot\! q)\left[\gamma_\mu^T(\gamma^T\!\cdot\! q)
      - (\gamma^T\!\cdot\! q)\gamma_\mu^T\right], \\
\tau^{7}_{1^-\mu} &= (q^T)^2\left(\gamma_\mu^T\,\gamma\!\cdot\! P
      - \gamma\!\cdot\! P\,\gamma_\mu^T\right)
      - 2q_\mu^T(\gamma^T\!\cdot\! q)\,\gamma\!\cdot\! P, \\
\tau^{8}_{1^-\mu} &= q_\mu^T(\gamma^T\!\cdot\! q)\,\gamma\!\cdot\! P,
\end{align}
where the transverse projection with respect to the total momentum $P$ is defined as
\begin{equation}
k_\mu^T = k_\mu - \frac{(P\!\cdot\!k)}{P^2} P_\mu.
\end{equation}
For AV mesons ($J^{PC}=1^{++},1^{+-}$), the same Lorentz structures are used,
with
\begin{equation}
\tau^{i}_{1^+\mu}(q,P)=\gamma_5\,\tau^{i}_{1^-\mu}(q,P),
\qquad i=1,\ldots,8.
\end{equation}

To simplify the angular dependence we expand the dressing functions in
Chebyshev polynomials of the second kind, $U_j(z)$, with
$z=\cos\theta=\hat q\cdot\hat P\in[-1,1]$, which satisfy
\begin{equation}
\int_{-1}^{1}\dd z\,\sqrt{1-z^2}\,U_m(z)U_n(z)
=
\frac{\pi}{2}\,\delta_{mn}.
\end{equation}

\section{Charge-conjugation constraints on axial--vector meson distribution amplitudes}
\label{app:Charge}

For a neutral AV channel, the BSA satisfies
\begin{equation}
C\,\Gamma_\nu^{T}(-k,P)\,C^{-1}
=
\xi_C\,\Gamma_\nu(k,P),
\label{eq:C_bsa}
\end{equation}
with
\begin{equation}
\xi_C=
\begin{cases}
+1, & a_1,\\
-1, & b_1.
\end{cases}
\label{eq:C_parity}
\end{equation}

We consider the zeroth moment of the longitudinal projection in Eq.~\eqref{eq:mellinVpar},
which can be written schematically as
\begin{equation}
\begin{aligned}
\langle x^0\rangle_\parallel^{1^+}
&\propto
\Tr_D \int^\Lambda \frac{\dd^4 k}{(2\pi)^4}\,
\gamma_5\,\gamma\!\cdot\!n\,n_\nu\\
&\quad\times
S_1(k_+)\,\Gamma_\nu(k,P)\,S_2(k_-).
\end{aligned}
\end{equation}
Applying charge conjugation to the RHS and using $\Tr M^T=\Tr M$, one obtains
\begin{equation}
\begin{aligned}
{\rm RHS}
\xrightarrow{C}\;
&\Tr_D \int^\Lambda \frac{\dd^4 k}{(2\pi)^4}\,
C\Big[
\gamma_5\,\gamma\!\cdot\!n\,n_\nu
\\
&\quad\times
S_1(k_+)\Gamma_\nu(k,P)S_2(k_-)
\Big]^T C^{-1}
\\
=
&\Tr_D \int^\Lambda \frac{\dd^4 k}{(2\pi)^4}\,
C\Big[
S_2^T(k_-)\Gamma_\nu^T(k,P)S_1^T(k_+)
\\
&\quad\times
n_\nu\,(\gamma\!\cdot\!n)^T\gamma_5^T
\Big] C^{-1}.
\end{aligned}
\label{eq:charge_step1}
\end{equation}
For the longitudinal projector, one finds
\begin{equation}
C\Big[(\gamma\!\cdot\!n)^T\gamma_5^T\Big]C^{-1}
=
(-\gamma\!\cdot\!n)\gamma_5
=
\gamma_5\,\gamma\!\cdot\!n.
\end{equation}
Meanwhile, the propagators transform as
\begin{equation}
C\,S_{1,2}^{T}(k_\mp)\,C^{-1}\;\to\; S_{1,2}(-k_\mp).
\end{equation}
Therefore,
\begin{equation}
\begin{aligned}
\mathrm{RHS}
\xrightarrow{C}\;
&\Tr_D \int^\Lambda \frac{\dd^4 k}{(2\pi)^4}\,
\xi_C\,\gamma_5\,\gamma\!\cdot\!n\,n_\nu
\\
&\times
S_2(-k_-)\,\Gamma_\nu(-k,P)\,S_1(-k_+).
\end{aligned}
\label{eq:B6}
\end{equation}
Under the change of integration variable $k\to -k$, one has
\begin{equation}
-k_- \to k_+,\qquad -k_+ \to k_-,
\end{equation}
so that, for $S_1=S_2$,
\begin{equation}
\begin{aligned}
&S_2(-k_-)\Gamma_\nu(-k,P)S_1(-k_+)
\\
&\longrightarrow
S_1(k_+)\Gamma_\nu(k,P)S_2(k_-).
\end{aligned}
\label{eq:charge_variable_change}
\end{equation}
One thus obtains
\begin{equation}
{\rm RHS}
\xrightarrow{C}
\xi_C\,\langle x^0\rangle_\parallel^{1^+}.
\end{equation}
Since the LHS is invariant under charge conjugation,
\begin{equation}
\langle x^0\rangle_\parallel^{1^+}
=
\xi_C\,\langle x^0\rangle_\parallel^{1^+}.
\end{equation}
Hence
\begin{equation}
\langle x^0\rangle_\parallel^{b_1}=0,
\end{equation}
whilst $\langle x^0\rangle_\parallel^{a_1}$ is left unconstrained by
charge conjugation.

Proceeding analogously for Eq.~\eqref{eq:mellinVperp}, one obtains the
corresponding constraint on the transverse moments,
\begin{equation}
\langle x^0\rangle_\perp^{1^+}
=
-\xi_C\,\langle x^0\rangle_\perp^{1^+},
\end{equation}
whence
\begin{equation}
\langle x^0\rangle_\perp^{a_1}=0,
\end{equation}
with $\langle x^0\rangle_\perp^{b_1}$ unconstrained.

The argument above uses $S_1=S_2$ at
Eq.~\eqref{eq:charge_variable_change}. When SU(3) flavour symmetry is
broken this step fails, the zeroth moments are no longer protected, and
their departure from zero measures the strength of the breaking. This
is the origin of the values
$\langle x^0\rangle^{\perpdir}_{K_1^{++}}=0.177$ and
$\langle x^0\rangle^{\para}_{K_1^{+-}}=0.289$ reported in
Table~\ref{tab:DA-moments-strange}.

\bibliographystyle{apsrev4-2}
\bibliography{bibliography}

\end{document}